\documentclass[prd,english,showkeys]{revtex4}%
\usepackage{amsmath}
\usepackage{graphicx}
\usepackage{epsfig}
\usepackage{dcolumn}
\usepackage{bm}
\usepackage{slashed}
\usepackage{amsfonts}
\usepackage{amssymb}
\usepackage{fontenc}
\usepackage{amsmath}
\usepackage{fontenc}
\usepackage{babel}%
\providecommand{\U}[1]{\protect\rule{.1in}{.1in}}
 
\begin{document}
\title{A note on "Measuring propagation speed of Coulomb fields" by R. de Sangro, G.
Finocchiaro, P. Patteri, M. Piccolo, G. Pizzella }
\author{D. M. Gitman$^{1,2,3}$}
\email{gitman@dfn.if.usp.br}
\author{A. E. Shabad$^{1,2}$}
\email{shabad@lpi.ru}
\author{A.A. Shishmarev $^{2,3}$}
\email{a.a.shishmarev@mail.ru}
\affiliation{$^{1}$\textsl{P. N. Lebedev Physics Institute, Leninsky Prospekt 53,
Moscow\ 117924, Russia} }
\affiliation{$^{2}$\textsl{Tomsk State University, Lenin Prospekt 36, Tomsk 634050, Russia} }
\affiliation{$^{3}$\textsl{Instituto de F\'{\i}sica, Universidade de S\~{a}o Paulo, CEP
05508-090, S\~{a}o Paulo, S. P., Brazil}}

\begin{abstract}
In connection with the discussion and the measurements fulfilled in Ref.
\cite{experiment}, the full identity is demonstrated between the Feynman
formula for the field of a moving charge and the Li\'{e}nard-Wiechert potentials.

\end{abstract}

\maketitle

In Ref. \cite{experiment} measurements were performed to decide between two
approaches to the field of a moving charge: one based on the
Li\'{e}nard-Wiechert potentials and the other on the Feynman interpretation.
The aim of the present note is to demonstrate that although apparently
different physical ideas are layed into the Feynman formula \cite{Feynman}, it
is as a matter of fact mathematically identical to that found in standard text
books, e. g. \cite{Landau2, Jackson}, for the Li\'{e}nard-Wiechert potentials
both for accelerated motion of the charge and its motion with a constant
speed. We believe that this observation should be taken into account and prove
to be useful in the general discussion on the matter \cite{Field}.

The Feynmann formula for the electric field $\mathbf{E}$\ of a moving
accelerated charge $q$\ is \cite{Feynman}
\begin{equation}
\mathbf{E}=\frac{q}{4\pi\epsilon_{0}}\left[  \frac{\mathbf{e}_{r^{\prime}}%
}{r^{\prime2}}+\frac{r^{\prime}}{c}\frac{d}{dt}\left(  \frac{\mathbf{e}%
_{r^{\prime}}}{r^{\prime2}}\right)  +\frac{1}{c^{2}}\frac{d^{2}}{dt^{2}%
}\mathbf{e}_{r^{\prime}}\right]  ,\label{A3_Feynman}%
\end{equation}
We set $c=1$ for the sake of simplicity and keep to the notations in
\cite{Feynman}. In (\ref{A3_Feynman}) the function%
\begin{equation}
r^{\prime}=\sqrt{x_{1}^{\prime}+x_{2}^{\prime2}+x_{3}^{\prime2}}\label{1}%
\end{equation}
is the modulus of vector $\mathbf{x}^{\prime},$ directed from the position
$\mathbf{\tilde{x}}$ of the moving charge to the observation point
$\mathbf{x}$:\
\begin{equation}
\mathbf{x}^{\prime}=\mathbf{x}-\mathbf{\tilde{x},}\label{2}%
\end{equation}
and $\mathbf{e}_{r^{\prime}}$ is the unit vector in the direction of
$\mathbf{x}^{\prime}.$\ The charge trajectory is given as $\mathbf{\tilde
{x}=\tilde{x}(}t^{\prime}),$ where $t^{\prime}$ is the time coordinate of the
charge. Therefore, with the location of the observation point fixed,
$\mathbf{x=const.,}$ we see that $\mathbf{x}^{\prime},$ as well as $r^{\prime
},$ is a function solely of $t^{\prime}.$ Once the influence of the charge
propagates exactly with the speed of light $c=1,$ the relation%
\begin{equation}
r^{\prime}(t^{\prime})=t-t^{\prime}\label{r'}%
\end{equation}
holds, where $t$ is the time of observation. With Eqs. (\ref{2}) and
(\ref{r'}) the length (\ref{1}) is just the distance between the position of
the charge and the observation point at the moment of emission.

Relation (\ref{r'}) defines $t$ as a function of $t^{\prime}.$Then, according
to the rule of differentiation of an inverse function, one has for any
function $a(t^{\prime})$%
\[
\frac{da\left(  t^{\prime}\right)  }{dt}=\frac{da\left(  t^{\prime}\right)
}{dt^{\prime}}\frac{dt^{\prime}}{dt}=\frac{da\left(  t^{\prime}\right)
}{dt^{\prime}}\left(  \frac{dt}{dt^{\prime}}\right)  ^{-1},
\]
where $\frac{dt}{dt^{\prime}}$ follows from (\ref{r'}) and (\ref{2}) to be%
\[
\frac{dt}{dt^{\prime}}=1+\frac{dr^{\prime}\left(  t^{\prime}\right)
}{dt^{\prime}}=1+\frac{d|\mathbf{x}-\mathbf{\tilde{x}}(t^{\prime}%
)|}{dt^{\prime}}=1-\frac{\left(  x_{i}-\tilde{x}_{i}(t^{\prime})\right)
}{r^{\prime}}\frac{d\tilde{x}_{i}\left(  t^{\prime}\right)  }{dt^{\prime}%
}=1-\frac{\left(  \mathbf{v\cdot x}^{\prime}\right)  }{r^{\prime}}=1-\left(
\mathbf{v\cdot e}_{r^{\prime}}\right)  .
\]
We have used here that $\frac{d\mathbf{\tilde{x}}\left(  t^{\prime}\right)
}{dt^{\prime}}=\mathbf{v}(t^{\prime})$ is the instantaneous speed of the charge.

Referring to the designation \ $\left(  \mathbf{e}_{r^{\prime}}\bar{v}\right)
=\kappa$ used for brevity we can now rewrite Eq. (\ref{A3_Feynman}) as
\begin{equation}
\mathbf{E}=\frac{q}{4\pi\epsilon_{0}}\left[  \frac{\mathbf{e}_{r^{\prime}}%
}{r^{\prime2}}+\frac{r^{\prime}}{\left(  1-\kappa\right)  }\frac{d}%
{dt^{\prime}}\left(  \frac{\mathbf{e}_{r^{\prime}}}{r^{\prime2}}\right)
+\frac{1}{\left(  1-\kappa\right)  }\frac{d}{dt^{\prime}}\left(  \frac
{1}{\left(  1-\kappa\right)  }\frac{d\mathbf{e}_{r^{\prime}}}{dt^{\prime}%
}\right)  \right]  . \label{A3_5}%
\end{equation}
Taking into account that
\begin{equation}
\frac{d\mathbf{e}_{r^{\prime}}}{dt^{\prime}}=\frac{d}{dt^{\prime}}\left(
\frac{\mathbf{x}^{\prime}}{r^{\prime}}\right)  =-\frac{\mathbf{v}}{r^{\prime}%
}+\frac{\kappa\mathbf{x}^{\prime}}{r^{\prime2}}=\frac{\kappa\mathbf{e}%
_{r^{\prime}}-\mathbf{v}}{r^{\prime}} \label{A3_6}%
\end{equation}
we calculate the second term in (\ref{A3_5}):%
\begin{equation}
\frac{r^{\prime}}{\left(  1-\kappa\right)  }\frac{d}{dt^{\prime}}\left(
\frac{\mathbf{e}_{r^{\prime}}}{r^{\prime2}}\right)  =\frac{3\kappa
\mathbf{e}_{r^{\prime}}-\mathbf{v}}{\left(  1-\kappa\right)  r^{\prime2}}.
\label{A3_7}%
\end{equation}
The third term in (\ref{A3_5}) is
\begin{equation}
\frac{1}{\left(  1-\kappa\right)  }\frac{d}{dt^{\prime}}\left(  \frac
{1}{\left(  1-\kappa\right)  }\frac{d\mathbf{e}_{r^{\prime}}}{dt^{\prime}%
}\right)  =\frac{1}{\left(  1-\kappa\right)  }\frac{d}{dt^{\prime}}\left(
\frac{\kappa\mathbf{e}_{r^{\prime}}-\mathbf{v}}{\left(  1-\kappa\right)
r^{\prime}}\right)  . \label{A3_8}%
\end{equation}
Let us calculate the derivative of $\left(  1-\kappa\right)  ^{-1}$:%
\begin{equation}
\frac{d\left(  1-\kappa\right)  ^{-1}}{dt^{\prime}}=\left(  1-\kappa\right)
^{-2}\left(  \frac{d\mathbf{e}_{r^{\prime}}}{dt^{\prime}}\cdot\mathbf{v}%
+\mathbf{e}_{r^{\prime}}\cdot\frac{d\mathbf{v}}{dt^{\prime}}\right)  =\left(
1-\kappa\right)  ^{-2}\left[  \frac{\kappa^{2}-v^{2}}{r^{\prime}}+\left(
\mathbf{e}_{r^{\prime}}\cdot\mathbf{v}\right)  \right]  , \label{A3_9}%
\end{equation}
where $\overset{\cdot}{\mathbf{v}}$ is the acceleration of the charge. Then
the third term in (\ref{A3_5}) becomes
\begin{align}
\frac{1}{\left(  1-\kappa\right)  }\frac{d}{dt^{\prime}}\left(  \frac
{1}{\left(  1-\kappa\right)  }\frac{d\mathbf{e}_{r^{\prime}}}{dt^{\prime}%
}\right)   &  =\frac{\left(  \kappa^{2}-v^{2}\right)  \left(  \kappa
\mathbf{e}_{r^{\prime}}-\mathbf{v}\right)  +\left(  1-\kappa\right)  \left[
2\kappa\left(  \kappa\mathbf{e}_{r^{\prime}}-\mathbf{v}\right)  +\mathbf{e}%
_{r^{\prime}}\left(  \kappa^{2}-v^{2}\right)  \right]  }{\left(
1-\kappa\right)  ^{3}r^{\prime2}}+\nonumber\\
&  +\frac{\left(  \kappa\mathbf{e}_{r^{\prime}}-\mathbf{v}\right)  \left(
\mathbf{e}_{r^{\prime}}\cdot\mathbf{v}\right)  +\left[  \mathbf{e}_{r^{\prime
}}\left(  \mathbf{e}_{r^{\prime}}\cdot\overset{\cdot}{\mathbf{v}}\right)
-\overset{\cdot}{\mathbf{v}}\right]  \left(  1-\kappa\right)  }{\left(
1-\kappa\right)  ^{3}r^{\prime}}. \label{A3_10}%
\end{align}
Finally, substituting (\ref{A3_7}) and (\ref{A3_10}) in (\ref{A3_5}) and
separating the factor $\left(  1-\kappa\right)  ^{3}r^{\prime2}$ , we get%
\begin{align}
\mathbf{E}  &  =\frac{q}{4\pi\epsilon_{0}\left(  1-\kappa\right)
^{3}r^{\prime2}}\left[  \mathbf{e}_{r^{\prime}}\left(  1-\kappa\right)
^{3}+\left(  3\kappa\mathbf{e}_{r^{\prime}}-\mathbf{v}\right)  \left(
1-\kappa\right)  ^{2}+\right. \nonumber\\
&  +\left(  \kappa^{2}-v^{2}\right)  \left(  \kappa\mathbf{e}_{r^{\prime}%
}-\mathbf{v}\right)  +2\kappa\left(  \kappa\mathbf{e}_{r^{\prime}}%
-\mathbf{v}\right)  \left(  1-\kappa\right)  +\mathbf{e}_{r^{\prime}}\left(
\kappa^{2}-v^{2}\right)  \left(  1-\kappa\right)  +\nonumber\\
&  +\left.  r^{\prime}\left(  \kappa\mathbf{e}_{r^{\prime}}-\mathbf{v}\right)
\left(  \mathbf{e}_{r^{\prime}}\cdot\overset{\cdot}{\mathbf{v}}\right)
+r^{\prime}\left[  \mathbf{e}_{r^{\prime}}\left(  \mathbf{e}_{r^{\prime}}%
\cdot\overset{\cdot}{\mathbf{v}}\right)  -\overset{\cdot}{\mathbf{v}}\right]
\left(  1-\kappa\right)  \right]  . \label{A3_11}%
\end{align}
It is easy to show that this is reduced to
\begin{align}
\mathbf{E}  &  =\frac{q}{4\pi\epsilon_{0}\left(  1-\kappa\right)
^{3}r^{\prime2}}\left[  \mathbf{e}_{r^{\prime}}\left(  1-v^{2}\right)
-\mathbf{v}\left(  1-v^{2}\right)  
\right.  \nonumber \\
& \left. +r^{\prime}\left(  \kappa\mathbf{e}%
_{r^{\prime}}-\mathbf{v}\right)  \left(  \mathbf{e}_{r^{\prime}}%
\cdot\overset{\cdot}{\mathbf{v}}\right)  +r^{\prime}\left(  \mathbf{e}%
_{r^{\prime}}\left(  \mathbf{e}_{r^{\prime}}\cdot\overset{\cdot}{\mathbf{v}%
}\right)  -\overset{\cdot}{\mathbf{v}}\right)  \left(  1-\kappa\right) \right] \nonumber \\
&  =\frac{q}{4\pi\epsilon_{0}}\frac{\left(  1-v^{2}\right)  \left(
\mathbf{x}^{\prime}-\mathbf{v}r^{\prime}\right)  }{\left(  r^{\prime
}-\mathbf{x}^{\prime}\cdot\mathbf{v}\right)  ^{3}}+\frac{q}{4\pi\epsilon_{0}%
}\frac{\left(  \mathbf{x}^{\prime}-\mathbf{v}r^{\prime}\right)  \left(
\mathbf{x}^{\prime}\cdot\overset{\cdot}{\mathbf{v}}\right)  -\overset{\cdot
}{\mathbf{v}}r^{\prime}\left(  r^{\prime}-\mathbf{x}^{\prime}\cdot
\mathbf{v}\right)  }{\left(  r^{\prime}-\mathbf{x}^{\prime}\cdot
\mathbf{v}\right)  ^{3}}. \label{A3_12}%
\end{align}
Taking into account that the numerator in the second term in the latter
expression can\ be rewritten as the double vector product%
\begin{equation}
\left(  \mathbf{x}^{\prime}-\mathbf{v}r^{\prime}\right)  \left(
\mathbf{x}^{\prime}\cdot\overset{\cdot}{\mathbf{v}}\right)  -\overset{\cdot
}{\mathbf{v}}r^{\prime}\left(  r^{\prime}-\mathbf{x}^{\prime}\cdot
\mathbf{v}\right)  =\left[  \mathbf{x}^{\prime}\times\left[  \left(
\mathbf{x}^{\prime}-r^{\prime}\mathbf{v}\right)  \times\overset{\cdot
}{\mathbf{v}}\right]  \right]  , \label{A3_13}%
\end{equation}
expression (\ref{A3_12}) can be recognized (with the identification
$q/4\pi\epsilon_{0}=e$, $\mathbf{x}^{\prime}=\mathbf{R}$\textbf{, }$r^{\prime
}=R$) as the expression ($63.8)$ for electric field in Ref. \cite{Landau2}.
Thus, expressions in Refs. \cite{Landau2} and \cite{Feynman} are the same.

\section*{Acknowledgements}

Supported by FAPESP under grants 2013/00840-9, 2013/16592-4 and 2014/08970-1,
by RFBR under Project 15-02-00293a, and by the TSU Competitiveness Improvement
Program, by a grant from \textquotedblleft The Tomsk State University D.I.
Mendeleev Foundation Program\textquotedblright. A.A.S. thanks also CAPES for
support. A.E.S. thanks the University of S\~{a}o Paulo for hospitality
extended to him during the period when this work was being performed.

\end{document}